\documentclass[9pt,conference]{IEEEtran}
\IEEEoverridecommandlockouts
\usepackage{cite}
\usepackage{amsmath,amssymb,amsfonts}

\usepackage{url}
\usepackage{booktabs}
\usepackage{algorithmic}
\usepackage{graphicx}
\usepackage{textcomp}
\usepackage{xcolor}
\usepackage{standalone}
\usepackage{float}
\usepackage{tikz}
\usepackage{orcidlink}

\usetikzlibrary {shapes.symbols} 
\usetikzlibrary{positioning}
\colorlet{mygreen}{green!60!black}
\colorlet{myred}{red!60!black}

\usepackage{cprotect}
\usepackage{siunitx}

\newcommand{\ctx}{\textit{context}}
\newcommand{\acc}{\textit{stem}}
\newcommand{\APA}{\ensuremath{\mathrm{APA}}}
\newcommand{\FAD}{\ensuremath{\mathrm{FAD}}}
\renewcommand{\bfseries}{\fontseries{b}\selectfont} 
\robustify\bfseries             
\newrobustcmd{\B}{\bfseries}    

\usepackage{pgfplots}
\usepackage{subcaption}  

\usepgfplotslibrary{colorbrewer}
\pgfplotsset{compat = 1.15, cycle list/Set1-8}
\usetikzlibrary{pgfplots.statistics, pgfplots.colorbrewer}
\usepackage{pgfplotstable}

\usepackage{csvsimple}

\usepackage{todonotes}

\begin{document}

\title{Accompaniment Prompt Adherence: A Measure for Evaluating Music Accompaniment Systems
}


\author{\IEEEauthorblockN{Maarten Grachten}
\IEEEauthorblockA{\textit{Sony Computer Science Laboratories} \\
  Paris, France \\
  \orcidlink{0000-0002-9488-0840} \url{https://orcid.org/0000-0002-9488-0840}
}
\and
\IEEEauthorblockN{Javier Nistal}
\IEEEauthorblockA{\textit{Sony Computer Science Laboratories} \\
Paris, France \\
javier.nistal@sony.com}
}

\maketitle

\begin{abstract}
Generative systems of musical accompaniments are rapidly growing, yet there are no standardized metrics to evaluate how well generations align with the conditional audio prompt.
We introduce a distribution-based measure called ``Accompaniment Prompt Adherence" (APA),
and validate it through objective experiments on synthetic data perturbations, and human listening tests.
Results show that APA aligns well with human judgments of adherence and is discriminative to transformations that degrade adherence.
We release a Python implementation of the metric using the widely adopted pre-trained CLAP embedding model, offering a valuable tool for evaluating and comparing accompaniment generation systems.
\end{abstract}


\section{Introduction}\label{sec:introduction}
AI-based music generation is rapidly evolving, with a significant focus on creating full musical mixes from text prompts~\cite{agostinelli2023musiclm,Forsgren_Martiros_2022_riffusion,schneider2023mousai,copet2023simple_musicgen}.
However, there is growing interest in models capable of generating an individual \emph{stem} (instrument/vocal part) to accompany an existing \emph{context} (a subset of the stems that make up a song) given as a prompt.
For example, given a context of drums and guitar, an accompaniment system may generate a bass stem to complement that context.
This task aligns more closely with music production workflows where artists iteratively add and refine stems to create a song~\cite{DBLP:conf/waspaa/Lattner19,grachten20:_bassn,wu2022jukedrummer,parker2024stemgen,pasini2024bass,nistal2024diff,mariani2023multi,tal2024joint_jasco}.\footnote{Note that the term accompaniment, as used here, is not strictly limited to the role of providing harmonic or rhythmic support, and may refer to any complementary musical part.}

Various metrics exist to evaluate different aspects of generated audio outputs.
The Fréchet Audio Distance (\FAD{})~\cite{kilgour2019frechet} is the most widely used for assessing audio quality, but doesn't specifically measure the output-to-prompt adherence of accompaniment generation systems.
Subjective evaluations are common for this task \cite{nistal2024diff,mariani2023multi,parker2024stemgen,tal2024joint_jasco}, but there is a lack of standardized objective metrics to measure how well the generated stem aligns with the context used as a prompt. 

In this work, we introduce a novel objective metric called Accompaniment Prompt Adherence (\APA{}), which uses a distribution-based embedding distance derived from \FAD{} to measure how well a \emph{candidate set} of context-stem pairs go together.
Based on \FAD{}, the \APA{} relies on a \emph{reference set} of context-stem pairs to evaluate the candidate set.
This means that, unlike other methods that rely on predefined notions of adherence or dedicated embedding models~\cite{parker2024stemgen,pasini2024bass,riou2024stem,ciranni2024cocola}, \APA{} is flexible, requires no training, and works with widely used embedding models like CLAP~\cite{wu23:_laionclap_large_scale_contr_languag_audio}.

We define \APA{} based on the intuition that adherence should be highest for matching context-stem pairs (i.e. from the same song and temporally aligned) and lowest for randomly assigned context-stem pairs across songs.
By measuring the effect of synthetic perturbations of the stems on the \APA{} scores, we 
 validate different configurations of the metric.
A comparison of the resulting optimal metric to human judgments of accompaniment adherence demonstrates a good fit. 
A Python package implementation is publicly released, providing a tool for evaluating music accompaniment generation systems.\footnote{\url{https://github.com/SonyCSLParis/audio-metrics}}

The paper is structured as follows: Sec.~\ref{sec:relwork} reviews existing methods for evaluating accompaniment generation systems.
Sec.~\ref{sec:method} introduces audio prompt adherence and details our experimental setup.
Sec.~\ref{sec:results} presents objective and subjective validation results.
Sec.~\ref{sec:discussion} provides a discussion, and Sec.~\ref{sec:conclusion} concludes with future research directions.

\section{Related Work}
\label{sec:relwork}
In the music accompaniment generation literature, most works include listening tests to assess how well the generated stems adheres to the given context.
Participants typically rate the quality or compatibility of a mix of the context and the generated stem using various methodologies \cite{parker2024stemgen,mariani2023multi,nistal2024diff,tal2024joint_jasco}.
While these subjective evaluations provide valuable insights, they can be time-consuming and are generally not available during training and experimentation.

To address these challenges, some works have introduced objective metrics to measure accompaniment adherence.
The MIRDD metric~\cite{parker2024stemgen} calculates the KL divergence between distributions of audio descriptors (such as pitch and rhythm) from mixes of the context with \emph{target} and \emph{generated} stems.
Other approaches involve accuracy metrics for melody, onset, and chords \cite{tal2024joint_jasco, wu2023music}.
However, these methods rely on fixed descriptors, making strong assumptions about what is relevant for measuring compatibility.

Another approach uses joint embedding models of context-stem compatibility \cite{pasini2024bass,riou2024stem,ciranni2024cocola,lattner2022samplematch}, 
defining a measure on the embeddings, like the CLAP score \cite{huang2023make} for text-audio compatibility.
While these approaches rely on training dedicated models, we propose a flexible, training-free method using pre-trained off-the-shelf embedding models like CLAP, making it more adaptable and relaxing the assumptions of descriptor-based or model-specific methods.

\section{Method}\label{sec:method}
\subsection{Preliminaries: Fréchet Audio Distance}

The Fréchet Audio Distance (\FAD{}) \cite{kilgour2019frechet} is a metric designed to evaluate the quality of synthesized audio by comparing it to real audio samples. Inspired by the Fréchet Inception Distance (FID) in image generation, \FAD{} compares the distributions of feature embeddings extracted from a pre-trained audio model for both a reference set (typically real audio examples) and a candidate set (typically generated samples). The distance between these two distributions is computed using the Fréchet distance, which measures the similarity of two multivariate Gaussians parameterized by their means and covariances.

Given two distributions \(\mathcal{R} \sim \mathcal{N}(\mu_r, \Sigma_r)\) (reference set) and \( \mathcal{C} \sim \mathcal{N}(\mu_c, \Sigma_c) \) (candidate set), the $\FAD_{C,R}$ is calculated as:
\[
\FAD_{C,R} = \lVert \mu_r - \mu_c \rVert^2 + \text{Tr}(\Sigma_r + \Sigma_c - 2(\Sigma_r \Sigma_c)^{1/2}),
\]

\noindent where \( \mu_r \) and \( \Sigma_r \) are the mean and covariance of the feature embeddings obtained from the reference set $R$ of audios, \( \mu_c \) and \( \Sigma_c \) are the mean and covariance of the feature embeddings from the candidate set $C$, \( \lVert \cdot \rVert \) denotes the Euclidean norm, and \( \text{Tr} \) is the trace of the matrix. 

The embeddings for \FAD{} are typically extracted from a pre-trained VGGish model, which is a variant of the VGG network trained on spectrogram representations of audio.
These features are used to calculate the means and covariances for the real and generated audio distributions, making \FAD{} a robust measure of both the fidelity and diversity of generated audio samples.
A recent comparison of different embedding models \cite{gui2023adapting} however, showed that \FAD{} is more effective for evaluating music generation systems when used in combination with embeddings from the CLAP model.

\subsection{Accompaniment Prompt Adherence}
\begin{figure}[t!]
    \centering
    \includegraphics[width=.8\columnwidth]{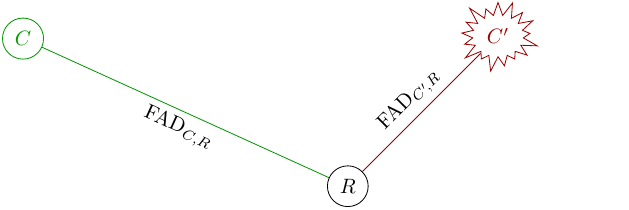}
    \includegraphics[width=.8\columnwidth]{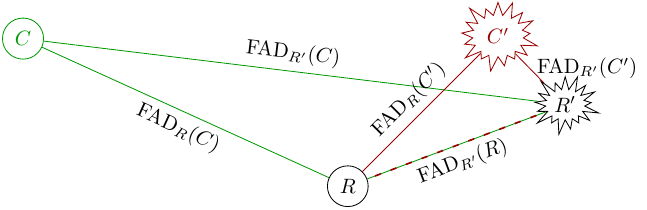}
    \caption{Absolute vs relative distances; Top: Counter-example for the naive approach, where a mismatched candidate set $C'$ is closer to (matched) reference set $R$ than the matched candidate set $C$ in absolute terms; Bottom: Given the same absolute distances, $C$ can still be closer to $R$, \emph{relative to} negative anchor $R'$}
    \label{fig:diagram}
\end{figure}

Given a reference set $R$ of matching context-stem pairs, a naive approach to measuring the prompt adherence of a candidate set of context-stem pairs $C$ is to downmix the pairs of $R$ and $C$ into single audio tracks respectively, and compute $\FAD_{C,R}$.
 The expectation here is that any musical incoherence in the context-stem pairs of the candidate set will cause a (proportional) shift in the embedding distribution with respect to that of the reference set, leading to a higher \FAD{} value.
 However, experiments using real music collections reveal that this is often not the case, especially when $R$ and $C$ are sampled from different music collections.
Figure~\ref{fig:diagram} (top) depicts this situation schematically, where $R$ is the reference set of matching context-stem pairs, $C$ is a candidate set of matching pairs, and $C'$ is obtained from $C$ by pairing contexts/stems at random. 

We postulate that a further set of mismatched pairs $R'$, constructed from $R$ by random pairing, will typically be closer to $C'$ than to $C$, as shown in Figure~\ref{fig:diagram} (bottom).
If true, then the relative proximity of the candidate set ($C$/$C'$) to $R$ and $R'$ is likely a good basis for measuring accompaniment prompt adherence.
Furthermore, as a metric, \FAD{} exhibits non-negativity and triangle inequality, implying that $|\FAD_{C,R'} - \FAD_{C,R}|$ is bounded by $\FAD_{R,R'}$ for any $C$.

We thus propose the following measure:

\begin{equation}
\small
\APA = \frac{1}{2} + \frac{\FAD_{C,R'} - \FAD_{C,R}}{2 \cdot \FAD_{R,R'}},
\label{eq:apa}
\end{equation}

\noindent which ranges from 0 to 1, where 1 indicates maximal adherence.
In practice, numerical instabilities occasionally lead to slight violations of the triangle inequality (yielding values outside [0, 1]).
We have not found this to pose a problem, and clip \APA{} to $[0, 1]$.
In the above definition, the \APA{} metric essentially measures the normalized difference between the distance of $C$ to $R'$ (low prompt adherence), ando to $R$ (high prompt adherence) respectively.

\subsection{Calculation pipeline}\label{sec:pipeline}
We calculate \APA{} on pairs of 5-second waveform windows, which provide enough context while remaining localized.
We transform the pairs into embedding vectors by first down-mixing the context and stem waveforms using various down-mixing regimes (Sec.~\ref{sec:mixing-cont-accomp}).
The down-mixed waveform is then passed through an embedding model (Sec.~\ref{sec:embedd-models}), producing one or more embedding vectors, which are averaged along the time dimension if necessary.
For some configurations, the embeddings are further projected onto a set of principal components computed from the reference set (Sec.~\ref{sec:pca-proj}).
The resulting embeddings are used to compute the \FAD{} scores for calculating \APA{} following Eq.~\ref{eq:apa}.

\subsubsection{mix regimes}\label{sec:mixing-cont-accomp}
The mix of \ctx{} and \acc{} waveforms is critical, as their relative levels significantly impact \FAD{} scores.
To judge the musical relationship between \ctx{} and \acc{}, they should ideally be equally audible.
However, since \ctx{} in our setup is typically a mix of multiple stems whereas \acc{} is a single one, loudness measures of each are not directly comparable, e.g., in a mix where \ctx{} and \acc{} are normalized to have equal integrated loudness, \acc{} tends to be dominant.
Thus, we evaluate different mix regimes based on both peak amplitude and integrated loudness measurements (see Tab.~\ref{tab:mix-regimes}).\footnote{We use \emph{pyloudnorm}~\cite{steinmetz2021pyloudnorm} for loudness normalization and \emph{cylimiter}~\cite{cylimiter} to prevent clipping}

\begin{table}[hbt!]
  \caption{Mix regimes used in the processing pipeline}
  \label{tab:mix-regimes}
  \fontsize{8}{9}\selectfont
  \centering
  \begin{tabular}{lp{.3\textwidth}c}
    \toprule
Type & Description    & Label \\
    \midrule
 peak & Preserve relative levels between \ctx{} and \acc{};
  normalize mix to original peak amplitude            & PP \\
 peak & Normalize both \ctx{} and \acc{} to -3\,dB       & P0 \\
 peak & Normalize \ctx{} to -3\,dB, \acc{} to -6\,dB        & P1 \\
 peak & Normalize \ctx{} to -3\,dB, \acc{} to -9\,dB        & P2 \\
 loudness & Normalize both \ctx{} and \acc{} to -20\,dB  & L0 \\
 loudness & Normalize \ctx{} to -20\,dB, \acc{} to -23\,dB  & L1 \\
 loudness & Normalize \ctx{} to -20\,dB, \acc{} to -26\,dB  & L2 \\
    \bottomrule
  \end{tabular}
\end{table}

\subsubsection{Embedders}\label{sec:embedd-models}
We compare three pre-trained models for extracting embeddings: VGGish
~\cite{hershey2017cnn}\cite{torchvggish}, OpenL3
~\cite{cramer2019look}\cite{openl3}, and CLAP
~\cite{wu23:_laionclap_large_scale_contr_languag_audio}.
VGGish (in the following denoted VGG), trained as an audio classifier on general audio data~\cite{abu-el-haija16:_youtube8m}, provides embeddings from its last feature layer (size 128) and has been widely used in music generation tasks despite not being specifically designed for musical characteristics.
OpenL3 (size 6144, denoted OL3) and CLAP, on the other hand, offer embeddings more tailored to musical nuances.
For CLAP, different pretrained models are available.
In this study, we evaluate the models trained on music and speech (CMS) and on music only (CM) \cite{clap-url}, using the last two feature layers (size 512 each) and output layer (size 128) of each model as embeddings (denoted CM(S)0/1/2 respectively).
By comparing these models, we aim to evaluate how well each embedding supports the \APA{} metric in representing acoustically and musically relevant features for measuring accompaniment adherence.

\subsubsection{Projections}\label{sec:pca-proj}
High-dimensional embedding spaces are more likely to be sparse, and may thus decrease the effectiveness of distribution-based metrics like \FAD, which measures the proximity of modes.
To address the varying dimensionalities of the embeddings, particularly the high dimensionality of OL3, we test whether lower-dimensional embeddings improve the effectiveness of FAD.
We consider two non-whitened\footnote{Preliminary experiments showed that whitening had a detrimental effect on the \APA{} values} PCA projections (PCA100 and PCA10) of the embeddings, as well the original embeddings (denoted NP).

\subsection{Music Collections}\label{sec:datasets}
This study uses five proprietary multitrack collections, and the publicly available MUSDB18~\cite{MUSDB18HQ}.
The genre distribution and size is not equal across collections.
MUSDB18 contains 150 Pop/Rock songs with an average of 4 stems per song, totaling approximately 9.8 hours of audio.
The largest proprietary collection features over 20,000 songs across various genres, with an average of 11-12 stems per song, amounting to 1,351 hours of audio. The smallest proprietary collection comprises 573 Trap sample packs, mainly short loops, with an average of 13 stems per segment and a total duration of 2.5 hours.
The other collections include Pop/Rock, and Production Music.
Each collection was split into non-overlapping, equal-sized reference and candidate datasets for the study.

\subsection{Validation Experiments}\label{sec:valid-exper}
We validate \APA{} using synthetic transformations on real data and conduct an ablation study of the calculation pipeline described in Sec.~\ref{sec:pipeline}. After identifying the best configuration, we perform subjective listening tests to compare \APA{} with human ratings on both real data and data generated by an existing accompaniment system.

\subsubsection{Objective Validation on Synthetic Data}\label{sec:experiment}
We begin by conducting an ablation study to determine the optimal setup for the \APA{} calculation pipeline.
Multiple candidate sets are derived from real $(\ctx, \acc)$ pairs by applying different transformations to \acc.
These transformations are grouped into two categories: those that should not affect \APA{} (labeled \textit{invariant}), such as adding noise or reconstructing the audio using a neural codec,\footnote{Invariance of the \APA{} measure to artifacts of such codecs is of special interest since they are often used in the generative models that the measure is intended for.} and those that should affect \APA{} (labeled \textit{non-invariant}), such as pitch and time-shifting.
Table~\ref{tab:transformations} lists all transformations and their naming convention. 

We rank the \APA{} calculation pipelines based on their ability to separate \emph{invariant} and \emph{non-invariant} transformations as measured by the Common Language Effect Size (CLES) \cite{mcgraw1992common}, the probability that an \APA{} value computed from an invariant transformation is higher than that of a non-invariant transformation.
We use 10,000 randomly sampled 5-second windows from reference and candidate sets to form $R$ and $C$, applying each of the eight transformations in Table~\ref{tab:transformations} to $C$ and then calculating \APA{} scores against the original $R$.

\begin{table}[hbt!]
\caption{Transformations of the \acc{} waveform used in the experiment.
Audio prompt adherence should be invariant to the upper four transformations, whereas should decrease by the lower four.}
\label{tab:transformations}
\fontsize{8}{9}\selectfont
\centering
\begin{tabular}{p{.28\textwidth}>{\centering}p{3em}c}
\toprule
Transformation of \acc & Invariant & Label \\
\midrule
Identity: original \acc   & Yes & TRUE \\
EnCodec \cite{defossez2022high} reconstruction & Yes & ENC \\
Descript \cite{kumar2024high} reconstruction & Yes & DAC\\
Add noise at original loudness - 20\,dB   & Yes & NOISE\\
\midrule
Time shift by 0.2 to 3.0s & No & TS \\
Pitch shift by +/- 1 to 7 semitones& No & PS \\
Time + Pitch shift & No & TPS \\
Randomly substitute \acc{} from other \ctx & No & SUBS \\
\bottomrule
\end{tabular}
\end{table}

\subsubsection{Subjective Evaluation}\label{sec:subj-eval}
To validate \APA{} against human ratings, we conduct a study of subjective audio prompt adherence.
Participants are presented with a 10-second music segment and asked to assess the compatibility of five different accompaniments, on a scale from 0 (no adherence) to 100 (perfect adherence), considering harmonic, rhythmic, and stylistic adherence aspects.
For the contexts we use examples from the largest music collection (see Section~\ref{sec:datasets}).
Among five candidate accompaniments presented to the listener for each context, we include the original ('Real'), one randomly chosen from the dataset ('Random'), and three produced by a generative model under different conditional settings ('Generated').

To generate data, we employ a recently proposed diffusion model~\cite{nistal2024diff} that generates instrumental stems given text and music audio context as input conditioning.
To present the context-stem pairs to the listener, we mix them slightly panned to left and right, respectively.
We perform loudness normalization of all individual audio segments to a loudness of -20\,dB LUFS.
A total of 875 ratings were obtained for a total of 100 segments.

\pgfplotsset{compat=1.17}
\pgfplotsset{/pgfplots/custom legend/.style={legend image code/.code={\draw [only marks,mark=square*]
plot coordinates {(0.cm,0cm)};}},}

\begin{center}
\begin{figure*}[hbt!]  
\hbox{
    \begin{subfigure}[t]{0.43\textwidth}  
        \captionsetup{margin={-7.5cm,0cm}}
        \caption{}
        \vspace{-0.5cm}
\begin{tikzpicture}
            \pgfplotstableread[col sep=comma]{cles2.csv}\csvdata
            \begin{axis}[
                axis x line*=bottom,
                axis y line*=left,
                enlarge y limits,
                width=0.8*\linewidth,
                xtick={1, 2, 3, 4, 5, 6, 7},
                xticklabels={PP,P0,P1,P2,L0,L1,L2},
                  ymin=0.2, ymax=0.8,
                ymajorgrids=true,
                ylabel={CLES},
            ]

            \addplot+[boxplot, fill=blue!25, draw=black, boxplot/draw direction=y] 
            table[x expr=1, y index=0] {\csvdata};

            \addplot+[boxplot, fill=blue!25, draw=black, boxplot/draw direction=y] 
            table[x expr=2, y index=1] {\csvdata};

            \addplot+[boxplot, fill=blue!25, draw=black, boxplot/draw direction=y] 
            table[x expr=3, y index=2] {\csvdata};

            \addplot+[boxplot, fill=blue!25, draw=black, boxplot/draw direction=y] 
            table[x expr=4, y index=3] {\csvdata};

            \addplot+[boxplot, fill=blue!25, draw=black, boxplot/draw direction=y] 
            table[x expr=6, y index=4] {\csvdata};

            \addplot+[boxplot, fill=blue!25, draw=black, boxplot/draw direction=y] 
            table[x expr=7, y index=5] {\csvdata};

            \addplot+[boxplot, fill=blue!25, draw=black, boxplot/draw direction=y] 
            table[x expr=8, y index=6] {\csvdata};
        \end{axis}
\end{tikzpicture}
        \label{fig:cles_mix}
    \end{subfigure}
    \hspace{-1cm}
    \begin{subfigure}[t]{0.43\textwidth}  
        \captionsetup{margin={-8.5cm,0cm}}
        \caption{}
        \vspace{-0.5cm}
\begin{tikzpicture}
            \pgfplotstableread[col sep=comma]{cles3.csv}\csvdata
            \begin{axis}[
                axis x line*=bottom,
                axis y line*=left,
                enlarge y limits,
                width=0.8*\linewidth,
                xtick={1,2,3,4,5,6,7,8},
                xticklabels={VGG,OL3,CMS0,CMS1,CMS2,CM0,CM1,CM2},
                xtick style={draw=none},
                ytick style={draw=none},
                ymajorticks=false,
                ymin=0.2, ymax=0.8,
                ymajorgrids=true,
                xticklabel style={rotate=-45},
                tick label style={font=\footnotesize}, 
            ]
            \addplot+[boxplot, fill=blue!25, draw=black, boxplot/draw direction=y] 
            table[x expr=1, y index=0] {\csvdata};

            \addplot+[boxplot, fill=blue!25, draw=black, boxplot/draw direction=y] 
            table[x expr=2, y index=1] {\csvdata};

            \addplot+[boxplot, fill=blue!25, draw=black, boxplot/draw direction=y] 
            table[x expr=3, y index=2] {\csvdata};

            \addplot+[boxplot, fill=blue!25, draw=black, boxplot/draw direction=y] 
            table[x expr=4, y index=3] {\csvdata};

            \addplot+[boxplot, fill=blue!25, draw=black, boxplot/draw direction=y] 
            table[x expr=6, y index=4] {\csvdata};

            \addplot+[boxplot, fill=blue!25, draw=black, boxplot/draw direction=y] 
            table[x expr=7, y index=5] {\csvdata};

            \addplot+[boxplot, fill=blue!25, draw=black, boxplot/draw direction=y] 
            table[x expr=8, y index=6] {\csvdata};

            \addplot+[boxplot, fill=blue!25, draw=black, boxplot/draw direction=y] 
            table[x expr=9, y index=7] {\csvdata};

        \end{axis}
\end{tikzpicture}
        \label{fig:cles_emb}

    \end{subfigure}    
    \hspace{-2cm}
    \begin{subfigure}[t]{0.43\textwidth}  
        \captionsetup{margin={-8.5cm,0cm}}
        \caption{}
        \vspace{-0.5cm}
        \raisebox{-5.3mm}{ 
\begin{tikzpicture}
 
            \pgfplotstableread[col sep=comma]{cles1.csv}\csvdata
            \begin{axis}[
                axis x line*=bottom,
                axis y line*=left,
                enlarge y limits,
                width=0.8*\linewidth,
                xtick={1, 2, 3},
                xticklabels={NP, PCA100, PCA10},
                xtick style={draw=none},
                ymin=0.2, ymax=0.8,
                ymajorgrids=true,
                ymajorticks=false
            ]

            \addplot+[boxplot, fill=blue!25, draw=black, boxplot/draw direction=y] 
            table[x expr=1, y index=0] {\csvdata};
            \addplot+[boxplot, fill=blue!25, draw=black, boxplot/draw direction=y] 
            table[x expr=2, y index=1] {\csvdata};
            \addplot+[boxplot, fill=blue!25, draw=black, boxplot/draw direction=y] 
            table[x expr=3, y index=2] {\csvdata};

        \end{axis}
\end{tikzpicture}
        \label{fig:cles_pca}
        }
    \end{subfigure}
}
    \vspace{0.1cm}
\hbox{
    \begin{subfigure}[t]{0.43\textwidth}  
        \captionsetup{margin={-7.5cm,0cm}}
        \caption{}
        \vspace{-0.5cm}
        \begin{tikzpicture}
        [thick, /pgfplots/boxplot/box extend=0.8,]
            \pgfplotstableread[col sep=comma]{intra.csv}\csvdata
            \begin{axis}[
                boxplot/draw direction=y,
                boxplot={
                        draw position={1/3 + floor(\plotnumofactualtype/4) + 1/3*mod(\plotnumofactualtype, 8)},
                        box extend=0.3,
                    },
                axis x line*=bottom,
                axis y line*=left,
                enlarge y limits,
                width=0.8*\linewidth,
                xtick={1, 3.3},
                xticklabels={Non-invariant, Invariant},
                xtick style={draw=none},
                ylabel={APA},
                ymin=0, ymax=1,
                custom legend,
                transpose legend,
                legend style={at={(0.05 ,1)}, font=\fontsize{7}{9}\selectfont,
                    anchor=north west,legend columns=4,
                    column sep=0.2em},
                legend cell align={left},
                ymajorgrids=true,
            ]

            \addplot+[boxplot, fill=blue!25, draw=black, boxplot/draw direction=y] 
            table[x expr=1, y index=0] {\csvdata};
            \addlegendentry{SUBS}
            
            \addplot+[boxplot, fill=blue!50, draw=black, boxplot/draw direction=y] 
            table[x expr=2, y index=1] {\csvdata};
            \addlegendentry{TPS}

            \addplot+[boxplot, fill=blue!75, draw=black, boxplot/draw direction=y] 
            table[x expr=3, y index=2] {\csvdata};
            \addlegendentry{PS}
            
            \addplot+[boxplot, fill=blue!100, draw=black, boxplot/draw direction=y] 
            table[x expr=4, y index=3] {\csvdata};
            \addlegendentry{TS}


            \addplot+[boxplot, fill=orange!25, draw=black, boxplot/draw direction=y] 
            table[x expr=6, y index=4] {\csvdata};
            \addlegendentry{NOISE}

            \addplot+[boxplot, fill=orange!50, draw=black, boxplot/draw direction=y] 
            table[x expr=7, y index=5] {\csvdata};
            \addlegendentry{ENC}
            
            \addplot+[boxplot, fill=orange!75, draw=black, boxplot/draw direction=y] 
            table[x expr=8, y index=6] {\csvdata};
            \addlegendentry{DAC}

            \addplot+[boxplot, fill=orange!100, draw=black, boxplot/draw direction=y] 
            table[x expr=9, y index=7] {\csvdata};
            \addlegendentry{TRUE}
    
            \end{axis}
        \end{tikzpicture}
        \label{fig:intra}
    \end{subfigure}
    \hspace{-1cm}
    \begin{subfigure}[t]{0.43\textwidth}  
        \captionsetup{margin={-8.5cm,0cm}}
        \caption{}
        \vspace{-0.5cm}
        \begin{tikzpicture}
        [thick, /pgfplots/boxplot/box extend=0.8,]
            \pgfplotstableread[col sep=comma]{inter.csv}\csvdata
            \begin{axis}[
                boxplot/draw direction=y,
                boxplot={
                        draw position={1/3 +                    floor(\plotnumofactualtype/4) + 
                        1/3*mod(\plotnumofactualtype, 8)},
                        box extend=0.3,
                    },
                axis x line*=bottom,
                axis y line*=left,
                enlarge y limits,
                width=0.8*\linewidth,
                xtick={1, 3.3},
                xticklabels={Non-invariant, Invariant},
                xtick style={draw=none},
                ymajorticks=false,
                ymin=0, ymax=1,
                ymajorgrids=true,
            ]

            \addplot+[boxplot, fill=blue!25, draw=black, boxplot/draw direction=y] 
            table[x expr=1, y index=0] {\csvdata};
            
            \addplot+[boxplot, fill=blue!50, draw=black, boxplot/draw direction=y] 
            table[x expr=2, y index=1] {\csvdata};

            \addplot+[boxplot, fill=blue!75, draw=black, boxplot/draw direction=y] 
            table[x expr=3, y index=2] {\csvdata};
            
            \addplot+[boxplot, fill=blue!100, draw=black, boxplot/draw direction=y] 
            table[x expr=4, y index=3] {\csvdata};


            \addplot+[boxplot, fill=orange!25, draw=black, boxplot/draw direction=y] 
            table[x expr=6, y index=4] {\csvdata};

            \addplot+[boxplot, fill=orange!50, draw=black, boxplot/draw direction=y] 
            table[x expr=7, y index=5] {\csvdata};
            
            \addplot+[boxplot, fill=orange!75, draw=black, boxplot/draw direction=y] 
            table[x expr=8, y index=6] {\csvdata};

            \addplot+[boxplot, fill=orange!100, draw=black, boxplot/draw direction=y] 
            table[x expr=9, y index=7] {\csvdata};
    
            \end{axis}
        \end{tikzpicture}
        \label{fig:inter}
    \end{subfigure}    
    \hspace{-1.9cm}
    \begin{subfigure}[t]{0.43\textwidth}  
        \captionsetup{margin={-8.5cm,0cm}}
        \caption{}
        \vspace{-0.5cm}
        \begin{tikzpicture}
[thick, /pgfplots/boxplot/box extend=0.8,]
    \pgfplotstableread[col sep=comma]{apa_sapa.csv}\csvdata
    \pgfplotstabletranspose\datatransposed{\csvdata}

    \begin{axis}[
        boxplot/draw direction=y,
        axis x line*=bottom,
        axis y line*=left,
        enlarge y limits,
        width=0.8*\linewidth,  
        xtick={1.5, 4.5, 7.5},
        xticklabel style={align=center, font=\small},
        xticklabels={Random, Generated, Real},
        xtick style={draw=none},
        ymin=0, ymax=1,
        custom legend,
        legend style={at={(0.05, 1)},font=\scriptsize,anchor=north west,legend columns=1,
        column sep=0.5em},
        ymajorgrids=true,
        ymajorticks=false
    ]

    \addplot+[boxplot, fill=blue!50, draw=black, boxplot/draw direction=y] 
    table[y index=1] {\datatransposed};
    \addlegendentry{APA}

    \addplot+[boxplot, fill=orange!70, draw=black, boxplot/draw direction=y] 
    table[y index=2] {\datatransposed};
    \addlegendentry{HUMAN}

    \addplot+[boxplot, fill=none, draw=none, opacity=0, boxplot/draw direction=y]{0};

    \addplot+[boxplot, fill=blue!50, draw=black, boxplot/draw direction=y] 
    table[y index=3] {\datatransposed};

    \addplot+[boxplot, fill=orange!70, draw=black, boxplot/draw direction=y] 
    table[y index=4] {\datatransposed};

    \addplot[boxplot, fill=none, draw=none, opacity=0, boxplot/draw direction=y]{0};

    \addplot+[boxplot, fill=blue!50, draw=black, boxplot/draw direction=y] 
    table[y index=5] {\datatransposed};

    \addplot+[boxplot, fill=orange!70, draw=black, boxplot/draw direction=y] 
    table[y index=6] {\datatransposed};

    \end{axis}
\end{tikzpicture}
        \label{fig:apasapa}
        
    \end{subfigure}
    }
    \caption{
      \emph{Top}: 
      CLES values (see Section~\ref{sec:valid-exper}) for (a) different mix regimes, (b) embedders , and (c) projections;
      \emph{Bottom}:
      The effect of invariant and non-invariant transformations on \APA{} values, for reference and candidate sets from (d) the \emph{same} music collection; and (e) \emph{different} collections; (f) Comparison of \APA{} values against subjective human ratings of accompaniment prompt adherence.
}

\end{figure*}
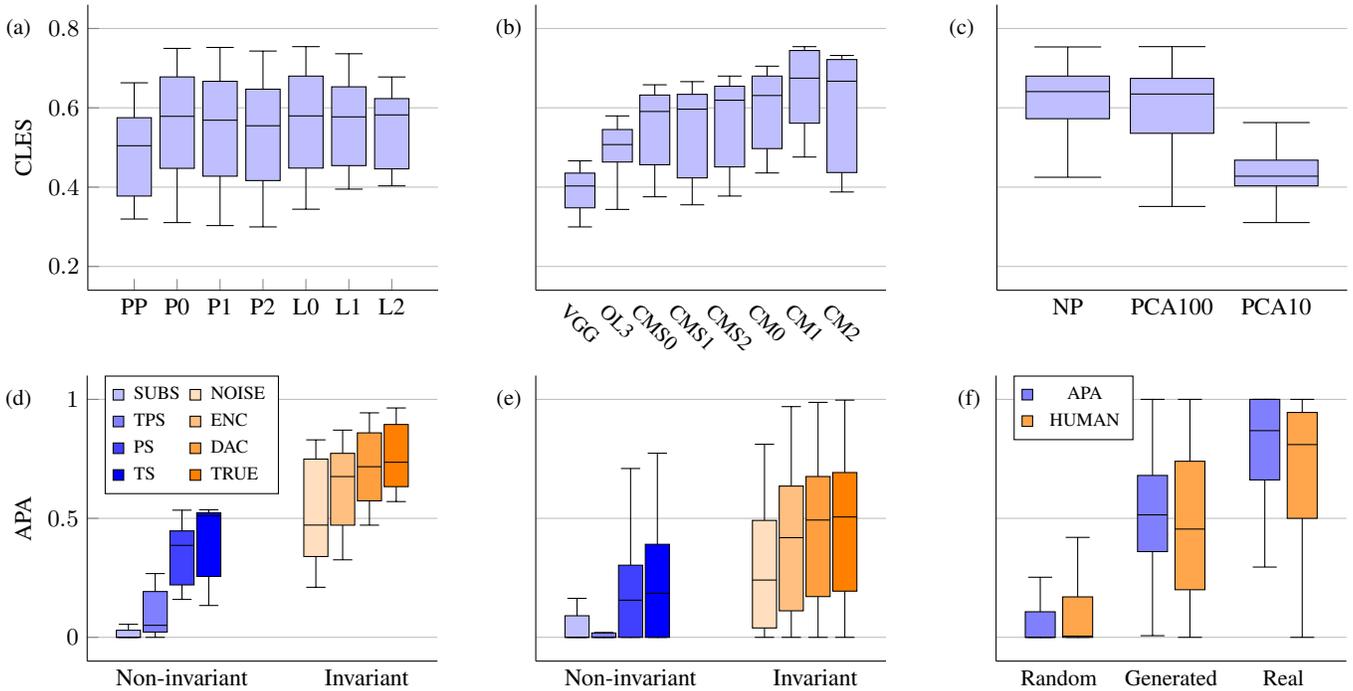
\end{center}

\section{Results}\label{sec:results}
As described in Section~\ref{sec:experiment}, we compare different configurations of \emph{mix regimes} (see Sec.\ref{sec:mixing-cont-accomp}), \emph{embedders} (see Sec.\ref{sec:embedd-models}) and \emph{projections} (see Sec.\ref{sec:pca-proj}) to calculate \APA{} and rank these based on CLES (Sec.~\ref{sec:experiment}).
We compute \APA{} values using all combinations of reference and candidate sets of the 6 collections (yielding 36 \APA{} values per combination).

Figures~\ref{fig:cles_mix},~\ref{fig:cles_emb}, and~\ref{fig:cles_pca} show the marginal distribution of CLES values for different mix regimes, embedders, and projections.
The CLAP embedder outperforms VGGish and OpenL3, in particular the last feature layer of the music only model (CM1).
On average, CLES values drop with increasing dimension reduction.
Finally, loudness-based mix regimes show less variance in CLES than peak-amplitude-based regimes.

Based on the CLES values, we select the best-performing configuration (L0, CM1, and PCA100) for further evaluation.
Figures~\ref{fig:intra} and \ref{fig:inter} show how the \APA{} score responds to various transformations as detailed in Table~\ref{tab:transformations}, grouped by the invariant/non-invariant categorization.
The results are shown separately for comparisons where the reference and candidate set are from the same collection (\ref{fig:intra}) and from different collections (\ref{fig:inter})
As expected, the TRUE and SUBS transformations define the upper and lower extremes of the \APA{} scale respectively, with the other transformations in between.
Neural codec artifacts do have a slight impact on \APA{} scores, more so for EnCodec (ENC) than for Descript (DAC).
Interestingly, adding white noise affects the scores substantially.
For intra-collection comparisons invariant and non-invariant transformations roughly populate the upper and lower half of the \APA{} scale, respectively.
\APA{} scores for inter-collection comparisons are generally lower, indicating distributional differences between collections.
Finally, time/pitch shifting significantly impacts on \APA{} scores, especially when combined.

Figure~\ref{fig:apasapa} compares \APA{} scores and \emph{human} ratings of audio prompt adherence, grouped by stem category.
For this comparison, the reference set used to compute the \APA{} scores for the rated examples consists of 50,000 5\,s windows randomly selected from the largest data collection.
For each stem category we randomly select 10 10s segments, and randomly sample 100 5s windows from them to obtain the candidate set.
We repeat this procedure 50 times, yielding 50 \APA{} values per stem category.
As expected, the user ratings are generally low for \emph{Random} and high for \emph{Real}, with \emph{Generated} in between.
This trend is clearly replicated in the \APA{} scores.

\section{Discussion}\label{sec:discussion}
The superiority of CLAP embeddings over VGGish and OpenL3 aligns with results from prior investigations on \FAD{} scores using different embedding models in the context of music \cite{gui2023adapting}.
The fact that CLAP was explicitly trained on music rather than general audio may be a clear advantage that pays off in downstream music tasks.

A notable result related to the mix regimes is that the original relative levels between \ctx{} and \acc{} (used in PP) are not optimal for computing \APA{} scores.
Instead, loudness normalization of \ctx{} and \acc{} (and subsequent application of an audio limiter) is more effective.
This is good because it means that there is no need to rely on user-set audio levels to compute reliable \APA{} scores.

Adding white noise to the stems had a substantial impact on the \APA{} scores.
An explanation for this may be that the addition of white noise reduces the distinction between matching and non-matching pairs---in the extreme case of an SNR of 0 there would objectively be no difference between matching and randomly paired stems and contexts, rendering the notion of prompt adherence meaningless.

The (moderate) invariance of \APA{} under neural audio codec reconstructions is a positive result since state-of-the-art generative models are likely to use such codecs.
The slight detrimental effect of the EnCodec reconstruction may be caused by the codec's characteristic artifacts that may occasionally alter pitch and timbre, which are relevant aspects in the perceived coherence of musical tracks.

Although the trend of inter-collection comparisons (Fig.~\ref{fig:inter}) mimics that of intra-collection comparisons results (Fig.~\ref{fig:intra}), it generally yields lower values and less contrast between invariant and non-invariant transformations, highlighting the importance of choosing a reference set that is representative of the context-stem pairs to be evaluated.
It also highlights the flexibility of \APA{} compared to metrics like MIRDD~\cite{parker2024stemgen}: rather than hard-coding assumptions of adherence as similarity of musical descriptors between context and stem, it relies on the reference set as implicitly defining accompaniment prompt adherence.
Note that the synthetic, transformation-based optimization of the \APA{} configuration tunes the definition toward a common sense (but ultimately subjective) notion of accompaniment prompt adherence, but does not encode this notion into the definition.

That the \APA{} metric as presented above aligns with human ratings of audio prompt adherence in the subjective listening test underscores its potential utility in the development, evaluation, and comparison of music accompaniment systems.
Furthermore, this result based on candidate sets of size 100 suggests that the \APA{} metric does not need large candidate samples to provide meaningful values, possibly allowing for per-song \APA{} scores, as was shown feasible for \FAD~\cite{gui2023adapting}.

\section{Conclusion}\label{sec:conclusion}
We proposed a novel measure for assessing audio prompt adherence, specifically designed to evaluate generative models of musical accompaniments.
Our approach formalizes adherence from a distributional perspective by evaluating the relative distances between candidate $(\ctx, \acc)$ pairs against reference sets of true pairs and randomly paired examples serving as positive and negative anchors.
Building on \FAD{}, we proposed a measure to assess audio prompt adherence quantitatively.
Our measure was tested through objective evaluations, comparing various calculation setups such as mix regimes, embedding models, and dimensionality reduction.
Using the optimal configuration, we validated the \APA{} measure through a subjective listening test, demonstrating its alignment with human judgments of audio prompt adherence.

Future work includes an investigation into the feasibility of per-song \APA{} values.
Furthermore, we plan to compare \APA{} to metrics based on dedicated models of accompaniment coherence \cite{riou2024stem,ciranni2024cocola}.

\clearpage

\bibliographystyle{IEEEtran}
\bibliography{icassp}

\end{document}